\documentclass[superscriptaddress,preprintnumbers,amssymb,nofootinbib]{revtex4-2}

\pdfoutput=1 

\usepackage[utf8x]{inputenc}
\usepackage{tikz-feynman}
\usepackage{subfig}
\usepackage{bm}
\usepackage{esvect}
\usepackage{verbatim}
\usepackage{multirow}

\pdfoutput=1
\usepackage[T1]{fontenc}
\usepackage{graphicx}
\usepackage{epsfig}
\usepackage{amsmath}
\usepackage{amsfonts}
\usepackage{amssymb}
\usepackage{braket}
\usepackage{cancel}
\usepackage{pstricks}
\usepackage{color}
\usepackage{feynmf}
\usepackage[normalem]{ulem}
\usepackage{epstopdf}

\def\lsim{\:\raisebox{-0.5ex}{$\stackrel{\textstyle<}{\sim}$}\:}
\def\gsim{\:\raisebox{-0.5ex}{$\stackrel{\textstyle>}{\sim}$}\:}

\usepackage{braket}
\usepackage{cancel}
\usepackage{setspace}
\usepackage{pstricks}
\usepackage{xcolor}
\usepackage{float}
\usepackage{mathtools}
\usepackage{booktabs}
\usepackage{enumitem}

\def\gsim{\lower0.5ex\hbox{$\:\buildrel >\over\sim\:$}}
\def\lsim{\lower0.5ex\hbox{$\:\buildrel <\over\sim\:$}}

\begin{document}

\title{Searching for New Physics with $b\bar{b} \ell^+ \ell^-$ Contact Interactions \\}

\author{Yoav Afik}
\email{yoavafik@campus.technion.ac.il}

\author{Shaouly Bar-Shalom}
\email{shaouly@physics.technion.ac.il}

\author{Jonathan Cohen}
\email{jcohen@campus.technion.ac.il}

\author{Yoram Rozen}
\email{rozen@physics.technion.ac.il}

\address{Physics Department, Technion--Institute of Technology, Haifa 3200003, Israel}

\date{\today}



\begin{abstract}
We study the impact of contact interactions involving two leptons (electrons or muons) and two $b$-quarks ($b \bar{b} \ell^+ \ell^-$) on the high-mass di-lepton region at the LHC.
We consider different selections of $b$-tagged jet multiplicities in the di-lepton final states: inclusive (no selection), 0, 1 and 2 $b$-tagged jets, and show that the single $b$-jet selection significantly improves the sensitivity to New Physics (NP) in the form of the $b \bar{b} \ell^+ \ell^-$ contact term.
We obtain a better sensitivity compared to the currently existing searches of NP in the di-lepton inclusive channel.
In particular, the expected limits go beyond competitive bounds set by LEP (for electrons) on the scale of NP, $\Lambda$, by a factor of $1.2-3.1$, depending on the chirality structure of the operator.
In addition, the expected limits on $\Lambda$, set by using a non-resonant LHC di-lepton inclusive search, are expected to be improved by a factor of $1.3-1.4$ for both electrons and muons.

\end{abstract}


\maketitle
\flushbottom

\section{Introduction}
\label{sec:intro}

The Standard Model (SM) of particle physics is believed to be a low-energy limit of a more fundamental high-energy theory.
The non-observability of direct NP signals at the Large Hadron Collider (LHC) pushes the scale of the underlying high-energy theory to the multi-TeV regime. 
Thus, within its energy and luminosity limitations, the NP is expected to show at the LHC in the tails of the distributions, where the NP signals can be modeled using Effective Field Theory (EFT) techniques, by integrating out the new degrees of freedom; this is in contrast to traditional NP searches of a resonance in the invariant mass spectrum, e.g. in case the new interactions originate from new heavy bosons. 

In this work we study the potential effects of NP in di-lepton signals at the LHC, assuming that the leading NP effects of the underlying heavy theory are in its interactions with the third generation quarks.
Indeed, the third generation quark doublet is naively expected to be the most sensitive to NP of a higher scale due to its relatively large mass, i.e., the masses of its quark components.
Furthermore, there have been some hints for NP involving the $b$-quark and violation of lepton flavor universality over the past several years, in measurements of $B$-meson decays~\cite{Aaij:2014pli,Aaij:2014ora,Aaij:2017vbb,Aaij:2015esa,Aaij:2015oid,Wehle:2016yoi,Abdesselam:2016llu,ATLAS:2017dlm,CMS:2017ivg,Bifani:2017gyn,Aaij:2019wad,Abdesselam:2019wac,Lees:2012xj,Lees:2013uzd,Huschle:2015rga,Hirose:2016wfn,Aaij:2015yra,Aaij:2017uff,Aaij:2017deq,Adamczyk:2019wyt,Abdesselam:2019dgh}. 
For example, the $R_{K^{(*)}}$ anomaly, which occurs in $b\to s \ell^+ \ell^-$ transitions where $\ell=\mu,e$, might very well imply a relatively low scale of NP which is intimately related to the third generation $b$-quark.
It is also worth mentioning that NP involving only the third generation quarks is less accessible (and thus less constrained) to high-energy processes at the LHC due to its significantly smaller content in the Parton Distribution Function (PDF).

We will consider a subset of dimension six 4-fermion operators involving a pair of $b$-quarks and a pair of charged leptons, which can be generated by a tree-level exchange of a heavy neutral vector boson ($Z'$).
Indeed, the striking possibility that the NP may lie at tree-level, has lead to extensive studies of specific model realizations, in particular, in conjunction with the observed $R_{K^{(*)}}$ anomaly, e.g., $Z'$ models~\cite{3rdgenZprimeBen,ZprimeRuderman,ZprimeKing,AssociatedZprime} and lepto-quarks~\cite{leptoquarks,LQ2}.
A more concerted effort has been put recently on the search for non-resonant deviations in the high mass regime of di-lepton events at the LHC (i.e., high invariant di-lepton mass $m_{\ell\ell}$) in the context of an EFT framework, placing bounds on higher dimensional 4-fermion operators of the form $q \bar{q} \ell^+ \ell^-$~\cite{Sirunyan:2018ipj,Chatrchyan:2012hda,Collaboration:2011tt,Aad:2011tq,Aad:2012bsa,Aad:2014wca,Aaboud:2016cth,ATLAS1,Ackerstaff:1996sf,Ackerstaff:1997nf,Abbiendi:1999wm,Abbiendi:2003dh,Barate:1999qx,Acciarri:2000uh,Schael:2006wu,Aaron:2011mv,Chekanov:2003pw,Abe:1991vn,Abe:1997gt,Abbott:1998rr,Abulencia:2006iv,Marzocca,SallyEFT,AngularEFT}. 
For example, $Z'$ gauge bosons which are heavy enough to evade traditional resonance searches~\cite{tradZprime} can still distort the shape of the $m_{\ell \ell}$ distribution~\cite{ZprimeRuderman}.

While these EFT studies treat the quarks uniformly, summing over all quark flavors and using the inclusive di-lepton sample to derive constraints on flavor-specific operators, we vary the $b$-tagged jet multiplicity of the di-lepton events and show that the exclusive 1 $b$-tagged jet selection can play a major role in deriving bounds on the strength of the $b \bar{b} \ell^+ \ell^-$ operators.
We have applied a similar approach in a previous work~\cite{bsll}, where we demonstrated the importance of an extra $b$-tagged jet selection for background rejection in studies of the NP effects due to a $b \bar{s} \ell^+ \ell^-$ contact term.
Other related studies include resonance type searches of associated $Z'$ and $W'$ production~\cite{AssociatedZprime,Wprime}, as well as direct searches for RPV SUSY in final states involving $b$-quarks~\cite{Soni}.

The paper is organized as follows: in Sec.~\ref{sec:model} we present the EFT framework and discuss the relevant constraints, in Sec.~\ref{sec:analysis} we present our proposed analysis in order to maximize the sensitivity, and in Sec.~\ref{sec:results} we present the expected results for an analysis at the LHC.
Finally, we conclude in Sec.~\ref{sec:conclusion}.

\section{Theoretical framework}
\label{sec:model}

Assuming that the NP is too heavy to be directly produced, we adopt an EFT approach where the heavy degrees of freedom of the underlying high-energy theory are integrated out.
In this case, the SM Lagrangian is augmented by a series of higher dimensional operators involving the interactions of the SM light fields and are suppressed by inverse powers of the NP scale $\Lambda$~\cite{EFT1,EFT2,EFT3,EFT4}.
In this work we will be interested in the subset of dimension six operators containing 4-fermion contact terms which involve the third generation $b$-quark and a pair of electrons and/or muons, that can be generated in the underlying theory by tree-level exchanges of a heavy vector boson.
Focusing on the $b$-quark interactions\footnote{Note that, due to gauge invariance, the operators involving the SU(2) lepton and quark doublets also contain 4-fermion contact interactions between the top-quark and the leptons as well as the charged interactions $(t \bar b)(e \bar\nu)$.
These operators are studied in Ref.~\cite{Afik:2020cvr}.}, it is convenient to cast the effective Lagrangian for the $b \bar{b} e^+ e^-$ and $b \bar{b} \mu^+ \mu^-$ terms in the form\footnote{As previously stated, the EFT approach in di-lepton signals at the LHC has been extensively utilized in the past, in particular, using 4-fermion contact operators~\cite{ATLAS1,Marzocca,SallyEFT,AngularEFT} and in $b$-decays~\cite{EFTbdecays1,EFTbdecays2,EFTbdecays3,StraubEFT}. 
In Ref.~\cite{Marzocca} it has been shown that 4-fermion operators of the form $q \bar{q} \ell^+ \ell^-$ can be probed via the high-$p_{T}$ di-muon tail, while others have highlighted the importance of angular variables in disentangling the NP effects~\cite{AngularEFT}.}: 
\begin{align}
\mathcal{L}_{eff} = \frac{g^2}{\Lambda^2}\sum_{i,j=L,R} \eta_{ij}(\bar{b}_{i} \gamma_{\mu} b_{i}) (\bar{\ell}_{j} \gamma^{\mu} \ell_{j}),
\label{eq:Lagrangian_bbll}
\end{align} 
where here $\ell=e,\mu$, $\Lambda$ is the NP scale and we have summed over all possible chirality structures with $\eta_{ij}=\pm1$, which will be useful for accounting for constructive/destructive interference with the SM Drell-Yan process $pp\to Z/\gamma^*\to\ell^+\ell^- + X$.
Also, we set $g=\sqrt{4\pi}$ following Ref.~\cite{Eichten:1983hw,Eichten:1984eu} and the analyses performed by the OPAL and ALEPH collaborations~\cite{Abbiendi:1998ea,Barate:1999qx,Schael:2006wu}.  
The choice $g=\sqrt{4\pi}$ for the coupling strength implies a strongly interacting underlying UV completion.
However, as previously stated, the effective interactions in Eq.~\eqref{eq:Lagrangian_bbll} naturally arise in the more broader EFT extension of the SM, the so-called SMEFT~\cite{EFT1,EFT2,EFT3,EFT4}.
In particular, the 4-fermion couplings defined in Eq.~\eqref{eq:Lagrangian_bbll} are linearly related to the Wilson coefficients of the corresponding gauge invariant operators in the SMEFT framework (see e.g.~\cite{Marzocca}) and we will, therefore, address below the sensitivity to the overall "effective" scale of NP $\Lambda/g$.

\begin{figure}[H]
  \centering
\begin{tikzpicture}
  \begin{feynman}
    \vertex (a10) {\(b\)};
    \vertex[below=1cm of a10] (a20);
    \vertex[below=1cm of a20] (a30) {\(\overline b\)};
    \vertex[right=1cm of a20] (b20);
    \vertex[right=2cm of a10] (c10) {\(\ell^+\)};
    \vertex[right=2cm of a30] (c30) {\(\ell^-\)};

    \vertex[above=0.1cm of b20] (l20) {\(\frac{g^{2}}{\Lambda^{2}}\)};

    \vertex[right=3cm of a10] (a11) {\(b\)};
    \vertex[below=1cm of a11] (a21);
    \vertex[below=1cm of a21] (a31) {\(g\)};
    \vertex[right=1cm of a21] (b21);
    \vertex[right=1cm of b21] (c21);
    \vertex[right=3cm of a11] (d11) {\(b\)};
    \vertex[below=1cm of d11] (d21) {\(\ell^+\)};
    \vertex[below=2cm of d11] (d31) {\(\ell^-\)};

    \vertex[above=0.1cm of c21] (l21) {\(\frac{g^{2}}{\Lambda^{2}}\)};

    \vertex[right=4cm of a11] (a12) {\(g\)};
    \vertex[right=1.5cm of a12] (a22);
    \vertex[right=1.5cm of a22] (a32) {\(b\)};
    \vertex[below=6em of a32] (b12) {\(\overline b\)};
    \vertex[left=1.5cm of b12] (b22);
    \vertex[left=1.5cm of b22] (b32) {\(g\)};
    \vertex[below=3em of a22] (c12);
    \vertex[right=0.5cm of c12] (c22);
    \vertex[below=0.5em of a22] (d12);
    \vertex[above=0.5em of b22] (d22);
    \vertex[right=1cm of c22] (e12);
    \vertex[right=0.5cm of e12] (e22);
    \vertex[above=0.5em of e22] (e32) {\(\ell^+ \)};;
    \vertex[below=0.5em of e22] (e42) {\(\ell^- \)};
    \vertex[left=0.1cm of c22] (l22) {\(\frac{g^{2}}{\Lambda^{2}}\)};

    \diagram* {
      {[edges=fermion]
        (d12) -- (a32),
        (b12) -- (d22),
      },
      (a10) -- [fermion] (b20),
      (a30) -- [anti fermion] (b20),
      (b20) -- [anti fermion] (c10),
      (b20) -- [fermion] (c30),
      
      (a11) -- [fermion] (b21),
      (a31) -- [gluon] (b21),
      (b21) -- [fermion] (c21),
      (c21) -- [fermion] (d11),
      (c21) -- [anti fermion] (d21),
      (c21) -- [fermion] (d31),

      (a12) -- [gluon] (d12),
      (d22) -- [gluon] (b32),
      (d12) -- [anti fermion] (c22),
      (d22) -- [fermion] (c22),
      (c22) -- [anti fermion] (e32),
      (c22) -- [fermion] (e42),

    };

  \end{feynman}
\end{tikzpicture}

	    \caption{Representative Feynman diagrams for a production of a lepton pair via the $b \bar{b} \ell^+ \ell^-$ operator at the LHC, in association with 0 (left), 1 (center) and 2 (right) $b$-jets. }
  \label{fig:Feynman_bbll}
\end{figure}
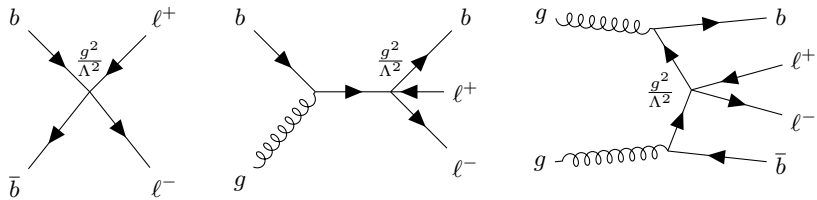

The operators in Eq.~\eqref{eq:Lagrangian_bbll} generate new tree-level di-lepton production modes at the LHC, in association with three different $b$-quark multiplicities, as depicted in Fig.~\ref{fig:Feynman_bbll}.
The behavior of the cross section for the corresponding $p p \to \ell^+ \ell^- + X$ processes can be parametrized by the contribution of three terms, which depend on the di-lepton invariant mass $m_{\ell \ell}$:
\begin{align}
\sigma(m_{\ell \ell})=\sigma^{SM}(m_{\ell \ell}) + \sigma^{NP}(g,\Lambda,m_{\ell \ell}) ~,
\label{eq:xsec_diff}
\end{align}
where $\sigma^{SM}$ is the pure SM term and $\sigma^{NP}(\Lambda,m_{\ell \ell})$ contain only the NP contributions: 
\begin{align}
\sigma^{NP}(g,\Lambda,m_{\ell \ell})=\frac{g^2}{\Lambda^2} \cdot \sigma^{SM \times NP}(m_{\ell \ell}) + \frac{g^4}{\Lambda^4} \cdot \sigma^{NP \times NP}(m_{\ell \ell}) ~,
\label{eq:xsec_sig}
\end{align}
where $\sigma^{SM \times NP}$ and $\sigma^{NP \times NP}$ contains the SM$\times$NP interference and NP$^2$ terms, respectively.
In particular, note that a negative $\sigma^{SM \times NP}$
corresponds to a destructive interference between the SM and the NP, so that  $\sigma^{NP}$
can have a negative value.
Furthermore, the contribution from these contact interactions grow with energy, i.e.,
$\sigma^{SM \times NP} \propto \hat s$ and $\sigma^{NP \times NP} \propto {\hat s}^2$, where $\sqrt{\hat s}$ is the center-of-mass energy of the hard process, and therefore a harder di-lepton spectrum is expected in events originating from these effective interactions. 
This requires a careful assessment of the EFT validity, which can be expressed by two expansion parameters:
\begin{align}
\mathcal{R}_{\Lambda/g} \equiv \frac{\hat s}{\Lambda^2/g^2} ~~,
~~ \mathcal{R}_{\Lambda} \equiv \frac{\hat s}{\Lambda^2} ~,\label{eq:EFTvalR}
\end{align}
that can be used to determine the EFT domain of applicability, as further explained below.

In particular, $\mathcal{R}_{\Lambda/g}$ is the EFT expansion parameter, i.e., the expansion of the effective Lagrangian at leading order in $g/\Lambda$. In our case, the expansion is up to dimension six operators: the SM$\times$NP interference term is of $\mathcal{O}({\cal R}_{\Lambda/g})$
while the NP$^2$ term is of $\mathcal{O}({\cal R}_{\Lambda/g}^2)$. Thus,
$\mathcal{R}_{\Lambda/g} < 1$ naively indicates the regime of validity of the EFT 
prescription. Moreover, the potential effects from operators of higher dimensions can be assessed from this expansion parameter. For example, the size of the contributions from the interference of additional dimension 8 operators with the SM is naively also of $\mathcal{O}({\cal R}_{\Lambda/g}^2)$. However, 
as will be explained below, this contribution from dimension 8 operators is strongly suppressed in our case, due to the rather high $m_{\ell\ell}$ selection we apply, which virtually eliminates any form of such interference effects with the SM.
Note also that, 
since the EFT framework is sensitive to the ratio $\Lambda/g$, its validity 
can be addressed by Eq.~\eqref{eq:EFTvalR} for the general coupling strength case and that a weaker coupling, e.g., $g\simeq\mathcal{O}(1)$, 
necessarily implies a weaker bound on $\Lambda$ (see Table \ref{tab:lumi_limits_bbll}).

On the other hand, the parameter $\mathcal{R}_{\Lambda}$ addresses the validity of the specific calculation within the EFT framework which may depend on the details of the underlying heavy physics and the process at hand. In particular, for an $s$-channel heavy NP exchange, the energy flow in the underlying scattering process may be estimated as $\hat s \simeq m^2_{\ell\ell}$, in which case $\mathcal{R}_{\Lambda} <1$ indicates the validity of the EFT 
prescription, since 
$m_{\ell\ell} < \Lambda$ ensures that the heavy excitation from the underlying NP 
cannot be produced on-shell. Note, though, that in some instances the 
EFT approach may hold even if $\mathcal{R}_{\Lambda} > 1$. 
For example, if the heavy NP is exchanged in the $t$-channel, then
the energy flow through the heavy propagator 
is effectively lower than $m_{\ell\ell}$, so that 
the validity of the EFT prescription also holds when $m_{\ell\ell} > \Lambda$. 
Therefore, the consistency of the calculation within the EFT framework as well as the legitimacy of the bounds on the heavy NP depends on 
how one interprets the details of underlying theory. As an example, 
a bound of $\Lambda/g > 3$ TeV which is obtained by examining 
an s-channel scattering process 
with $\sqrt{\hat s} = 4$ TeV is not valid from the EFT point of view if $g=1$ (since in this case the value of the bound is smaller than the energy flow in the process), but it is legitimate if $g=2$ in the underlying heavy theory. We will further address this point below when discussing our results.

%

Let us now briefly discuss the relevant constraints on our setup.
It is useful to cast those in terms of a third generation $Z'$ (i.e., a $Z'$ which couples dominantly to the third generation quarks) which can UV-complete our EFT description in Eq.~\eqref{eq:Lagrangian_bbll}.
In that case the $Z'$ mass is associated with the scale of NP up to some coupling, i.e., $\Lambda\simeq M_{Z'}$ and $g^2\simeq g_{bb} g_{\ell \ell}$.
Direct constraints on the flavor-diagonal $b \bar{b} e^+ e^-$ interaction were derived by the OPAL and ALEPH collaborations at LEP~\cite{Ackerstaff:1997nf,Abbiendi:1998ea,Barate:1999qx,Schael:2006wu} and are given in Tab.~\ref{tab:bbll_limits_bbll}, where the operators were normalized as in Eq.~\eqref{eq:Lagrangian_bbll}.  
Similar bounds were extracted recently on the $b\bar b\mu^+\mu^-$ operator in~\cite{Marzocca} from non-resonant LHC di-lepton searches.
Constraints on the $b \bar{b} \mu^+ \mu^-$ operators from lepton flavor universality tests in $\Upsilon$-meson turn out to be very weak as they only apply to the low-mass region~\cite{ZprimeKing}.
One can also use the bounds on the flavor non-diagonal $b \bar{s} \ell^+ \ell^-$ operator to constrain the flavor diagonal $b \bar{b} \ell^+ \ell^-$ ones, if the $g_{bb}$ coupling is assumed to be related to the $b-s$ admixture via $g_{bs}=V_{ts}g_{bb}$, where $V_{ts} \sim 0.04$  is the $t-s$ CKM element~\cite{Tanabashi:2018oca}.
In particular, it has been shown that the most stringent constraint on such $Z'$ vector bosons comes from the low energy $\mathcal{B}_s -\bar{\mathcal{B}}_s$ mixing and reads $g_{bs}\lesssim M_{Z'}/194\,\text{TeV}$~\cite{3rdgenZprimeBen}.
On the other hand, the $Z'$ coupling to muons, $g_{\mu\mu}$, which for left-handed muons need not be too large so as to evade the bound from neutrino trident $\nu_{\mu}\mu^+\mu^-$ production~\cite{ZprimeKing}, induces subdominant constraints from $Z\to 4\mu$ at the LHC which only excludes the low-mass region $5\lesssim M_{Z'}\lesssim 70\,\text{GeV}$, whereas the di-muon resonance search in ATLAS~\cite{ATLAS2} constrains $Z'$ masses of up to 5 TeV. 

Henceforth, we will only refer to the LEP and the non-resonant LHC searches for comparison, since those directly bound the diagonal contact interactions under consideration.

\section{Analysis}
\label{sec:analysis}

\subsection{Simulated Event Samples}

Monte Carlo (MC) simulated event samples of $pp$ collisions at $\sqrt{s} = 13$ TeV were used to estimate the SM contribution as well as the EFT signal.
All MC samples have been generated using a minimal requirement for $m_{\ell \ell}$, in order to have high statistics in the tail of the $m_{\ell \ell}$ distribution.
Furthermore, all of the samples were generated at Leading Order (LO) using {\sc MadGraph5\_aMC@NLO 2.6.3}~\cite{Alwall:2011uj}  in the 5 flavor scheme with the NNPDF30LO PDF set~\cite{Ball:2014uwa} and interfaced with the {\sc Pythia 8.23}~\cite{Mrenna:2016sih} parton shower.
The default {\sc MadGraph5\_aMC@NLO} LO dynamical scale was used, which is the transverse mass calculalted by a $k_T$-clustering of the final-state partons.
Events of different jet-multiplicities are matched using the MLM scheme~\cite{MLM}.
All simulated samples were processed through {\sc Delphes 3}~\cite{deFavereau:2013fsa} in order to simulate the detector effects and apply simplified reconstruction algorithms.

Background processes with $t \bar{t}$ events, as well as DY processes ($Z/ \gamma^* + jets$), which are the dominant SM backgrounds, were generated with up to two extra partons.
Other backgrounds which were taken into account are $t \bar{t}+W$, $t \bar{t}+Z$ (denoted in the text as "top" together with $t \bar{t}$ events) and $VV$ processes ($WW,WZ,ZZ$). 
For simplicity, events with fake electrons from $W + jets$ and multi-jet processes were neglected, as those are expected to be sub-dominant~\cite{ATLAS1}.

A valid {\sc FeynRules} UFO model~\cite{Degrande:2011ua} was built in order to generate signal events, using the EFT prescription presented in Sec.~\ref{sec:model}.
The signal was generated with up to two extra partons, taking into account both the interference of the NP with the SM and the pure NP contribution, i.e. only the terms in Eq.~\eqref{eq:xsec_sig}.
While only discrete values of $\Lambda$ were generated for all possible chirality combinations, the yields of non discrete values of $\Lambda$ were derived by means of interpolation.

\subsection{Event Reconstruction}

As noted, in this analysis, a reconstruction of electrons, muons and hadronic jets is used with {\sc Delphes 3}~\cite{deFavereau:2013fsa}.
For leptons (electrons and muons), reconstruction is applied based on truth-level leptons by applying transverse momentum ($p_{\mathrm{T}}$)- and pseudo-rapidity ($\eta$)-dependent artificial efficiency weight.
An isolation from other energy-flow objects is applied in a cone of $R=0.4$, and a minimum $p_{\mathrm{T}}$ requirement of $20$ GeV for each lepton.
Jets were reconstructed using the anti-$k_{t}$~\cite{Cacciari:2008gp} clustering algorithm with radius parameter of $R=0.4$ implemented in FastJet~\cite{Cacciari:2011ma,Cacciari:2005hq}, and are required to have transverse momentum $p _ {\mathrm{T} } >20$ GeV and pseudo-rapidity $\left|\eta\right|<2.5$.
The identification of $b$-tagged jets was done by applying a $p_ {\mathrm{T}}$-dependent weight based on the jet's associated flavor, using truth information, and the MV2c20 tagging algorithm~\cite{ATL-PHYS-PUB-2015-022} in the 70\% working point.
The $c$-jet and $light$-jet rejections for this working point are 8.1 and 440, respectively.

\subsection{Event Selection}

As a base-point, the selection contains two leptons, either electrons or muons, with Opposite-Sign (OS) charges.
The invariant mass of both leptons ($m_{\ell \ell}$) was used for optimization; the NP is expected to dominate at the tail of the $m_{\ell \ell}$ distribution whereas a small yield for the background is expected in that regime.
Different selections of $b$-tagged jet multiplicities in each event were tested: inclusive (no selection), 0, 1 and 2 $b$-tagged jets.
The integrated luminosity was chosen to be 140 fb$^{-1}$, which is the approximate full LHC Run-2 integrated luminosity.

An optimization was done by maximizing the sensitivity of the selection.
The sensitivity was estimated by using the expected $Z$-value calculated by the \verb|BinomialExpZ| function by \verb|RooFit|~\cite{Verkerke:2003ir}.
The expected $Z$-value is defined as the number of standard deviations from the background-only hypothesis, given a signal yield and background uncertainty.
In one of the recent analyses by the ATLAS collaboration~\cite{ATLAS1}, the relative background uncertainty for an inclusive selection was 8\% for both final states (di-electron or di-muon) with $m_{\ell \ell} > 1200$ GeV.
For a tighter selection of $m_{\ell \ell} > 1800$ GeV, the relative background uncertainty was estimated to be 16\% (13\%) for di-electron (di-muon) final states.
Similarly, the relative signal uncertainty for an inclusive selection was 8-9\% (12-13\%) for di-electron (di-muon) final states with $m_{\ell \ell} > 1200$ GeV.
For a tighter selection of $m_{\ell \ell} > 1800$ GeV, the relative signal uncertainty was estimated to be 9-10\% (13-14\%) for di-electron (di-muon) final states.
Uncertainties related to the selection of $b$-tagged jets are expected to be small.
See for example Ref.~\cite{Sirunyan:2020lgh}, where a measurement of final states with leptons and $b$-tagged jets is performed by the CMS collaboration with a jet flavor tagging uncertainty to be in the order of 1\%.
Guided by those analyses, we take the total relative uncertainty on the background to be 25\%, which is on the conservative side and we assume that the signal uncertainty is included within this selection.

\begin{figure}[H]
\centering
\includegraphics[width=0.49\textwidth]{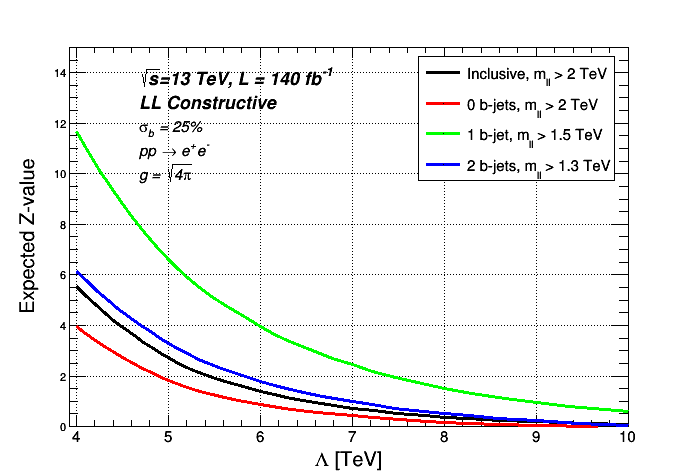}
\includegraphics[width=0.49\textwidth]{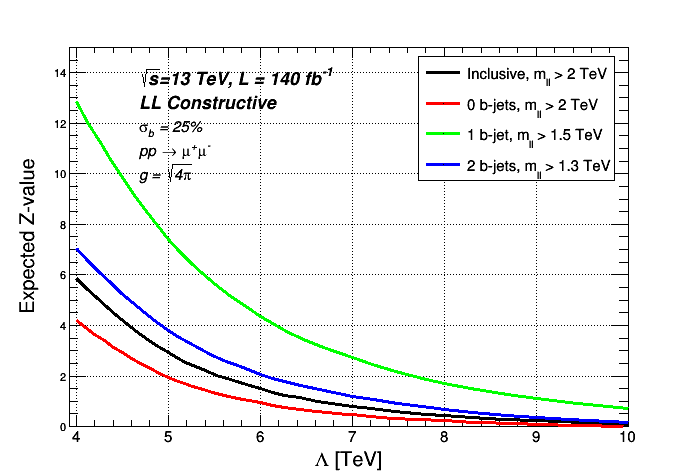}
\caption{Expected $Z$-value for different signal hypotheses varied with respect to the NP scale value ($\Lambda$), for the selections of number of $b$-tagged jets discussed in the text, for electrons (left) and muons (right).
}
\centering
\label{fig:Z_value}
\end{figure}

\begin{figure}[H]
\centering
\includegraphics[width=0.49\textwidth]{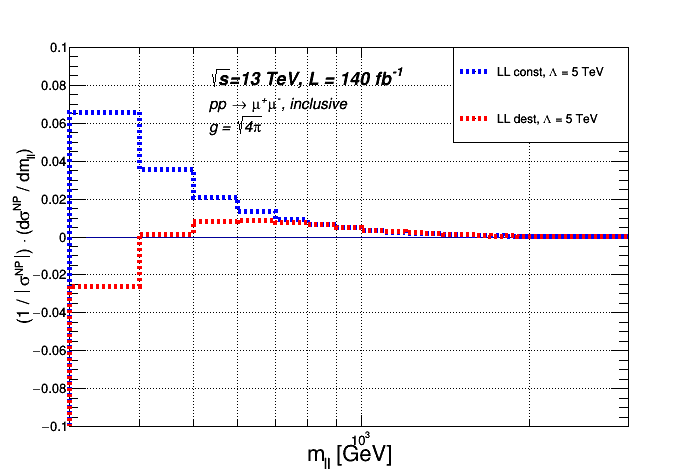}
\includegraphics[width=0.49\textwidth]{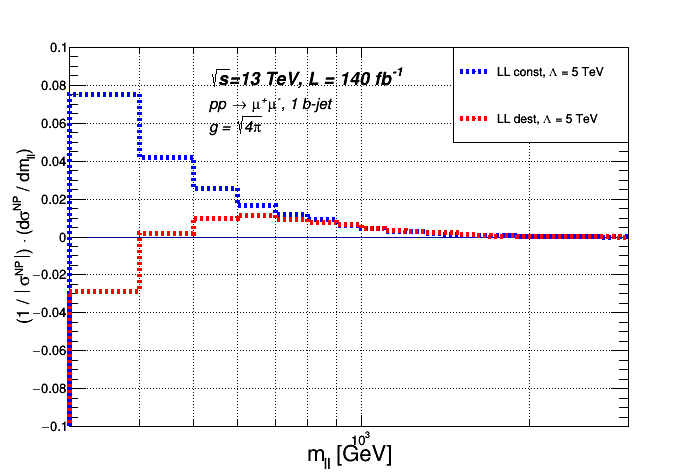}
\caption{
Normalized differential NP cross section with respect to the di-lepton invariant mass, $\frac{1}{\left|\sigma^{NP}\right|} \frac{d\sigma^{NP}}{dm_{\ell \ell}}$, corresponding {\it only} to $\sigma^{NP}$ in Eq.~\eqref{eq:xsec_sig} (i.e. {\it only} to the SM$\times$NP interference and the NP$^2$ terms).
One constructive ($\eta_{LL}=+1$) and one destructive ($\eta_{LL}=-1$) $b \bar{b} \ell^+ \ell^-$ operator are shown, with $g=\sqrt{4\pi}$ and $\Lambda = 5$ TeV. An inclusive selection is presented at the left plot, and a single $b$-tagged jet selection is presented at the right plot.
Final states with muons are presented.
}
\centering
\label{fig:inclusive_1b_normalized}
\end{figure}

We find that the selection with the highest sensitivity is the one with a single $b$-tagged jet, as can be observed from Fig.~\ref{fig:Z_value}. 
Furthermore, the expected sensitivity is higher for final states with muons compared to final states with electrons. 
The reason for that is a higher detector acceptance in the $m_{\ell \ell}$ tail for the di-muon final states, as we get by using the {\sc Delphes 3} detector simulation.
For example, for the $LL$ operator with $\Lambda = 5$ TeV and a constructive interference, in the 1 $b$-tagged jet selection and $m_{\ell \ell} > 1500$ GeV, the expected signal yield is 25 events for the muon operators and 16 events for the electron ones, while the corresponding cross sections are similar.
The background for final states with muons comparing to electrons is higher as well for similar reasons.
However, the sensitivity for the muons final state is still higher when combining both of the signal and background yields.
In Fig.~\ref{fig:inclusive_1b_normalized} we present
the normalized $m_{\ell \ell}$ signal distributions, for the two selections of lepton pair: an inclusive and one $b$-tagged jet selections. 
Looking at the signal with the destructive interference, an interesting behavior is observed: for low values of $m_{\ell \ell}$, the interference between the SM and NP processes is more dominant than the pure NP contribution, leading to a negative contribution of the signal to the nominal SM $m_{\ell \ell}$ distribution.
At high values of $m_{\ell \ell}$, however, the pure NP contribution dominates while the interference contribution is negligible, due to the fact that for high $m_{\ell \ell}$ selections, which are sufficiently higher than the $Z$-boson mass, the relative SM contribution is considerably suppressed (for example, the interference term is 
more than 2(3) orders of magnitudes smaller 
at $\Lambda \sim 5(10)$ TeV, with the selection of $m_{\ell \ell} > 1.5$ TeV).

\begin{figure}[H]
\centering
\includegraphics[width=0.49\textwidth]{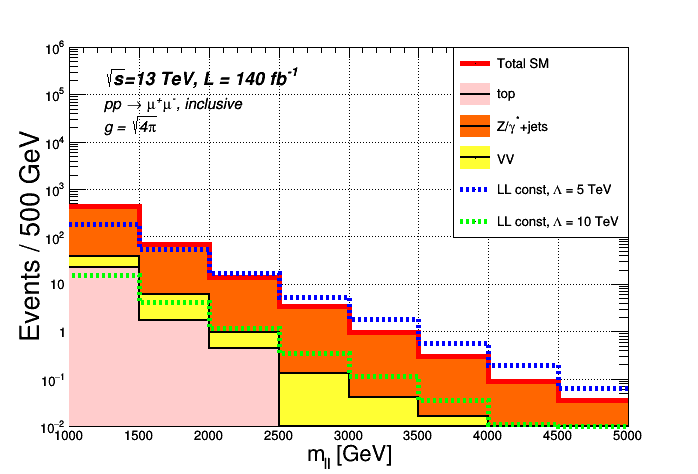}
\includegraphics[width=0.49\textwidth]{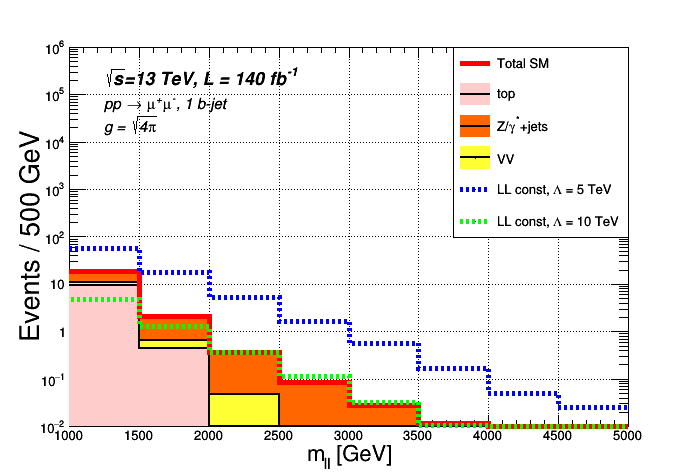}
\caption{
Distribution of the invariant mass of both leptons 
for two benchmark signal scenarios, with $\eta_{LL}=+1$ and $\Lambda = 5,10$ TeV, overlaid with the SM stacked background processes.
An inclusive selection is presented in the left plot, and a single $b$-tagged jet selection is presented in the right plot.
Final states with muons are presented.
}
\centering
\label{fig:inclusive_1b_observables}
\end{figure}

In Fig.~\ref{fig:inclusive_1b_observables} we present the $m_{\ell \ell}$ event distributions for the signal and for the SM background, for the two selections of lepton pair: an inclusive and one $b$-tagged jet selections.
For both inclusive and 0 $b$-tagged jet selections, the maximum sensitivity is found to be with $m_{\ell \ell} > 2$ TeV.
On the other hand, for the 1 $b$-tagged jet selection, a value of $m_{\ell \ell} > 1.5$ TeV gives maximum sensitivity, whereas for the 2 $b$-tagged jets selection maximum sensitivity is obtained for $m_{\ell \ell} > 1.3$ TeV.
For all selections, only events with $m_{\ell \ell} < 5$ TeV are considered.
The maximum sensitivity selections were found to be common between final states with electrons and muons.
The $m_{\ell \ell}$ selection for all types of the number of $b$-tagged jet selections are summarized in Tab.~\ref{tab:bbll_mumu_selections}.
As mentioned earlier, with this high invariant mass selection the interference between the SM and the NP contribution is negligible.

\begin{table}[htp]
\begin{center}
\begin{tabular}{| l | c | c | c | c |}
\hline
Observable & inclusive & 0 $b$-tag & 1 $b$-tag & 2 $b$-tag   \\
\hline
$\mathcal{N}_{b}$ & - & =0 & =1 & =2 \\
$\mathcal{N}_{\mu} / \mathcal{N}_{e}$ & = 2, OS & = 2, OS & = 2, OS & = 2, OS \\
$m_{\ell \ell}$ [TeV] & $> 2$ & $> 2$ & $> 1.5$ & $> 1.3$ \\
\hline
\end{tabular}
\end{center}
  \caption{Summary of the maximum sensitivity ranges for the various selections of lepton pairs in association with $b$-tagged jets.
  }
\label{tab:bbll_mumu_selections}
\end{table}

\subsection{Ratio Analysis - a note}
In case where the NP couples only to one lepton generation, either with constructive or destructive interference, a ratio analysis may be more effective for detecting a deviation from the SM prediction.
In this case, the theoretical uncertainties can be minimized since they are expected to be similar in both the di-muon and di-electron final states, therefore reducing the total relative uncertainty.
This is important in particular in the high energy regime of $m_{\ell \ell}$ where one expects a substantial theoretical uncertainty. Such a ratio analysis is in particular useful for the study of lepton flavor non-universality, and will be investigated in detail in Ref.~\cite{Afik:2020cvr}, within a broader context including contact interactions involving the top-quark with the leptons.
The use of ratio observables in lepton non-universality 
studies has been prevalent, e.g., in di-lepton production at the LHC \cite{Marzocca}, 
in top-quark decays \cite{Kamenik:2018nxv} and in B-decays 
from both the theoretical 
\cite{ratioth1,ratioth2,ratioth3,ratioth4,ratioth5,ratioth6} and 
experimental \cite{ratioexrev,Aaij:2015yra,Aaij:2017deq,Aaij:2017uff} standpoints.

\section{Results}
\label{sec:results}

As can be seen from Fig.~\ref{fig:Z_value}, a higher sensitivity for the 1 $b$-tagged jet selection is expected in comparison with the other selections.
The 2 $b$-tagged jet selection yields better sensitivity than the inclusive and 0 $b$-tagged jet selections, but still not compatible with the 1 $b$-tagged jet selection.
The reason the 1 $b$-tagged jet selection possesses the best sensitivity is that the $Z/ \gamma^* + jets$ and $VV$ backgrounds dominate mostly the inclusive and the 0 $b$-tagged jet selections, so that the requirement to have a single $b$-tagged jet removes most of those backgrounds.
For the 2 $b$-tagged jet selection the $Z/ \gamma^* + jets$ and $VV$ backgrounds are reduced as well, however, the signal yield is lower compared to the 1 $b$-tagged jet selection.

\begin{figure}[H]
\centering
\includegraphics[width=0.49\textwidth]{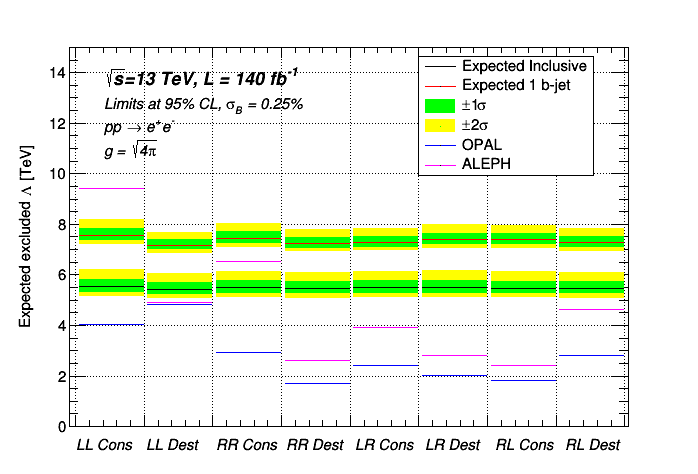}
\includegraphics[width=0.49\textwidth]{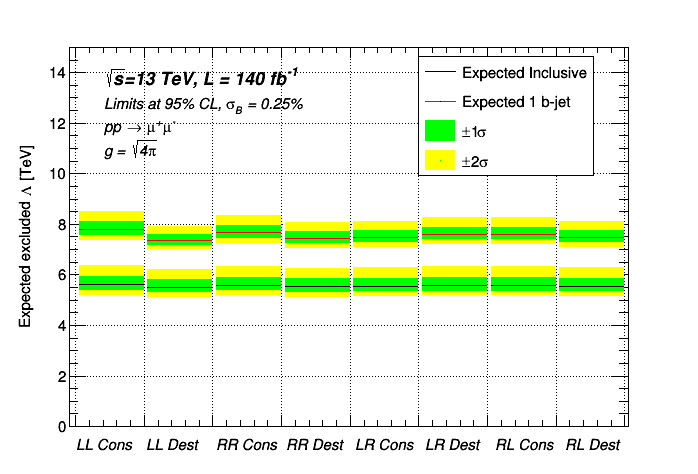}
\caption{
Expected upper limit on $\Lambda$ for all possible chirality structures of the $b \bar{b} \ell^+ \ell^-$ operator. 
The cases where no $b$-jets requirement are used (inclusive, lower bands) and exactly one $b$-tagged jet is required (upper bands) are presented for comparison, all with 25\% background uncertainty. Left: $b \bar{b} e^+ e^-$ operators; right: $b \bar{b} \mu^+ \mu^-$ operators.
The bounds from LEP in the di-electron case are also presented for comparison.
}
\centering
\label{fig:Lambdas}
\end{figure}

In order to determine the sensitivity to the NP scale $\Lambda$, we calculated the $p$-value for each signal and background hypothesis using the
\verb|BinomialExpP| function by \verb|RooFit|~\cite{Verkerke:2003ir}.
After the $p$-value of the background-only and background+signal hypotheses for each point were calculated, a $CL_{s}$~\cite{Read:2002hq} test was made in order to determine whether the corresponding signal point is expected to be excluded with 95\% Confidence Level (CL).
The expected upper limits for the inclusive and for the single $b$-tagged jet selection for all possible chirality structures of the $b \bar{b} \ell^+ \ell^-$ interactions are presented in Fig.~\ref{fig:Lambdas}, where the current operating luminosity at the LHC was used. 
The $\pm 1 \sigma$ ($\pm 2 \sigma$) bands are derived by calculating the limits after pulling the background up and down with the corresponding background uncertainty, which we define as 25\% for $1 \sigma$ (50\% for $2 \sigma$).
The bounds from LEP in the di-electron case are also plotted in Fig.~\ref{fig:Lambdas} for comparison.
We find, for example, that the sensitivity of the exclusive 1 $b$-tagged jet selection can extend the reach of this operator up to $\Lambda = 7.6^{+0.2}_{-0.2}\,(7.8^{+0.2}_{-0.3})$ TeV for electrons (muons) for the $LL$ constructive operator.
The sensitivity to the NP scale $\Lambda$ for the operators with all types of chiralities are summarized in Tab.~\ref{tab:bbll_limits_bbll}, where we also included the LEP bounds for comparison. 
The limits on similar operators which were obtained in~\cite{Marzocca}, using an inclusive di-lepton ATLAS analysis~\cite{ATLAS1}, are not presented, since the authors of this work used a $2\sigma$ method, while we choose a more conservative approach - $CL_{s}$.
Thus, although the results of both works are not comparable, it is clear that better limits will be obtained by using the single $b$-tagged jet selection, regardless of the statistical method.

\begin{table}[htp]
\begin{center}
\begin{tabular}{| c | c | c | c | c | c | c | c | c | c |}
\hline
Operator & Channel & $LL$ const & $LL$ dest & $RR$ const & $RR$ dest & $LR$ const & $LR$ dest & $RL$ const & $RL$ dest \\
\hline
\multirow{6}{*}{$b\bar{b} e^+ e^-$} & OPAL [TeV] & 4.0 & 4.8 & 2.9 & 1.7 & 2.4 & 2.0 & 1.8 & 2.8 \\
& ALEPH [TeV] & 9.4 & 4.9 & 6.5 & 2.6 & 3.9 & 2.8 & 2.4 & 4.6 \\
& Inclusive [TeV] & $5.5^{+0.2}_{-0.3}$& $5.4^{+0.2}_{-0.3}$ & $5.5^{+0.2}_{-0.3}$ & $5.5^{+0.2}_{-0.3}$ & $5.5^{+0.2}_{-0.3}$ & $5.5^{+0.2}_{-0.3}$ & $5.5^{+0.2}_{-0.3}$ & $5.5^{+0.2}_{-0.3}$ \\
& 0 $b$-tag [TeV] & $4.9^{+0.2}_{-0.2}$ & $4.9^{+0.2}_{-0.2}$ & $4.9^{+0.2}_{-0.2}$ & $4.9^{+0.2}_{-0.2}$ & $4.9^{+0.2}_{-0.2}$ & $4.9^{+0.2}_{-0.3}$ & $4.9^{+0.2}_{-0.2}$ & $4.9^{+0.2}_{-0.2}$ \\
& 1 $b$-tag [TeV] & $7.6^{+0.2}_{-0.2}$ & $7.2^{+0.2}_{-0.2}$ & $7.5^{+0.2}_{-0.2}$ & $7.2^{+0.2}_{-0.2}$ & $7.3^{+0.2}_{-0.2}$ & $7.4^{+0.2}_{-0.2}$ & $7.4^{+0.2}_{-0.2}$ & $7.3^{+0.2}_{-0.2}$ \\
& 2 $b$-tag [TeV] & $6.0^{+0.1}_{-0.2}$ & $5.7^{+0.1}_{-0.2}$ & $5.8^{+0.1}_{-0.2}$ & $5.7^{+0.1}_{-0.2}$ & $5.8^{+0.1}_{-0.2}$ & $5.8^{+0.1}_{-0.2}$ & $5.8^{+0.1}_{-0.2}$ & $5.8^{+0.1}_{-0.2}$ \\
\hline
\multirow{4}{*}{$b\bar{b} \mu^+ \mu^-$} & Inclusive [TeV] & $5.6^{+0.2}_{-0.3}$ & $5.5^{+0.2}_{-0.3}$ & $5.6^{+0.2}_{-0.3}$ & $5.5^{+0.2}_{-0.3}$ & $5.6^{+0.2}_{-0.3}$ & $5.6^{+0.2}_{-0.3}$ & $5.6^{+0.2}_{-0.3}$ & $5.6^{+0.2}_{-0.3}$ \\
& 0 $b$-tag [TeV] & $5.0^{+0.2}_{-0.3}$ & $4.9^{+0.2}_{-0.3}$ & $5.0^{+0.2}_{-0.3}$ & $5.0^{+0.2}_{-0.3}$ & $5.0^{+0.2}_{-0.3}$ & $5.0^{+0.2}_{-0.3}$ & $5.0^{+0.2}_{-0.3}$ & $5.0^{+0.2}_{-0.3}$ \\
& 1 $b$-tag [TeV] & $7.8^{+0.2}_{-0.3}$ & $7.3^{+0.2}_{-0.2}$ & $7.7^{+0.2}_{-0.3}$ & $7.4^{+0.2}_{-0.3}$ & $7.5^{+0.2}_{-0.3}$ & $7.6^{+0.2}_{-0.3}$ & $7.6^{+0.2}_{-0.3}$ & $7.5^{+0.2}_{-0.3}$ \\
& 2 $b$-tag [TeV] & $6.2^{+0.2}_{-0.2}$ & $5.9^{+0.1}_{-0.2}$ & $6.1^{+0.2}_{-0.2}$ & $6.0^{+0.1}_{-0.2}$ & $6.0^{+0.1}_{-0.2}$ & $6.1^{+0.2}_{-0.2}$ & $6.1^{+0.2}_{-0.2}$ & $6.0^{+0.2}_{-0.2}$ \\

\hline
\end{tabular}
\end{center}
  \caption{Summary of the limits on the scale of NP, $\Lambda$, assuming $g=\sqrt{4\pi}$, for the $b \bar{b} e^+ e^-$ and $b \bar{b} \mu^+ \mu^-$ operators for constructive ($\eta_{ij}=+1$) and destructive ($\eta_{ij}=-1$) interference with the SM.
  The bounds from LEP in the di-electron case are also presented for comparison.
  }
\label{tab:bbll_limits_bbll}
\end{table}

\begin{figure}[H]
\centering
\includegraphics[width=.49\textwidth]{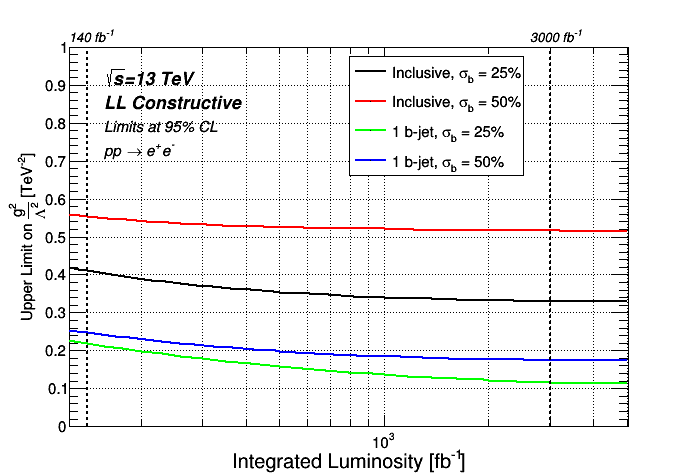}
\includegraphics[width=.49\textwidth]{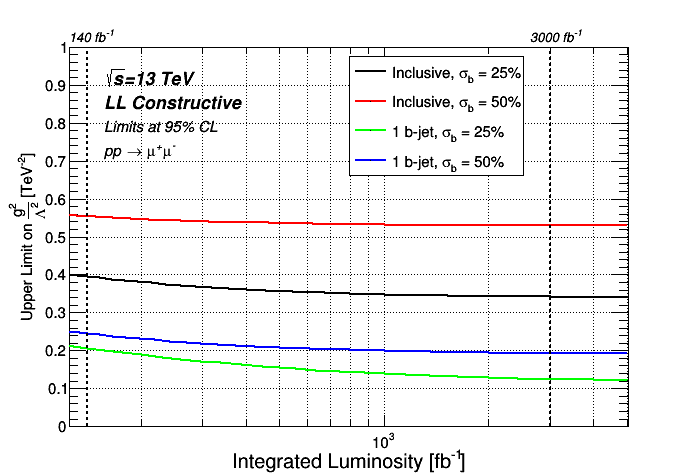}
\caption{
Expected upper limit on $g^2 / \Lambda^2$ for final states with $b$-quarks and electrons (left) or muons (right), as a function of the total integrated luminosity, for operators with the $LL$ chirality that have constructive interference with the SM ($\eta_{LL}=+1$).
The cases where no $b$-jets requirement is used (inclusive) and exactly one $b$-tagged jet is required are presented for comparison, with 25\% and 50\% background uncertainty.
}
\centering
\label{fig:Lumi}
\end{figure}

In Fig.~\ref{fig:Lumi} we furthermore show, for the $LL$ constructive operator, the expected upper limit on $g^2 / \Lambda^2$, for the inclusive and for the 1 $b$-tagged jet selections and for different values of the total integrated luminosity, using two scenarios of the relative background uncertainty: 25\% and 50\%\footnote{We note that the limits can be slightly improved for different values of the integrated luminosity by a dedicated optimization.
For simplicity, we keep the selections used by the optimization we did for 140 fb$^{-1}$.}.
The expected limits on $\Lambda/g$ 
from Fig.~\ref{fig:Lumi} are listed in Tab.~\ref{tab:lumi_limits_bbll} for two values of the integrated luminosity: 140 fb$^{-1}$ and 3000 fb$^{-1}$.
Evidently, even in cases where the relative uncertainty of the 1 $b$-tag selection is significantly higher compared to the inclusive selection, better sensitivity can still be obtained using this channel.
Also, an increase of the luminosity from 140 fb$^{-1}$ to 3000 fb$^{-1}$ improves the expected limits on $g^2 / \Lambda^2$ by 24\% (16\%) for electrons (muons) with an inclusive selection, and by 87\% (66\%) for electrons (muons) with a single $b$-tagged jet selection, assuming 25\% of background uncertainty.

\begin{table}[htp]
\begin{center}
\begin{tabular}{| c | c | c | c | c | c |}
\hline
Operator & Integrated luminosity & Inclusive, 25\% Unc. & Inclusive, 50\% Unc. & 1 $b$-tag, 25\% Unc. & 1 $b$-tag, 50\% Unc. \\
\hline
\multirow{2}{*}{$b\bar{b} e^+ e^-$} & 140 fb$^{-1}$ & $\Lambda / g > 1.6$ TeV & $\Lambda / g > 1.3$ TeV & $\Lambda / g > 2.1$ TeV & $\Lambda / g > 2.0$ TeV \\
& 3000 fb$^{-1}$ & $\Lambda / g > 1.7$ TeV & $\Lambda / g > 1.4$ TeV & $\Lambda / g > 2.9$ TeV & $\Lambda / g > 2.4$ TeV \\
\hline
\multirow{2}{*}{$b\bar{b} \mu^+ \mu^-$} & 140 fb$^{-1}$ & $\Lambda / g > 1.6$ TeV & $\Lambda / g > 1.3$ TeV & $\Lambda / g > 2.2$ TeV & $\Lambda / g > 2.0$ TeV \\
& 3000 fb$^{-1}$ & $\Lambda / g > 1.7$ TeV & $\Lambda / g > 1.4$ TeV & $\Lambda / g > 2.8$ TeV & $\Lambda / g > 2.3$ TeV \\
\hline
\end{tabular}
\end{center}
  \caption{Summary of the limits on the scale of NP scaled by the corresponding coupling, $\Lambda / g$, for the $b\bar{b} e^+ e^-$ and $b\bar{b} \mu^+ \mu^-$ $LL$ operators with constructive interference ($\eta_{ij}=+1$). The expected limits for two benchmark values of the integrated luminosity are presented.
  }
\label{tab:lumi_limits_bbll}
\end{table}

Finally, let us discuss the consistency of the bounds in Table~\ref{tab:lumi_limits_bbll} with the EFT approach, following our discussion 
in Section \ref{sec:model} about the EFT validity criteria outlined 
in Eq.~\eqref{eq:EFTvalR}. For example, the bounds on $\Lambda$ obtained 
with the 1 $b$-tagged jet selection and the 25\% uncertainty scenario 
are decreased from $\Lambda\sim{7-8}$ TeV in the case of $g=\sqrt{4\pi}$, to $\Lambda\sim{2-3}$ TeV for $g=1$, i.e., for a "natural" weakly coupled underlying NP.
The former result for the strong coupling case is clearly consistent with the EFT criteria $\mathcal{R}_{\Lambda} < 1$ (see discussion below Eq.~\eqref{eq:EFTvalR}), since we have applied an overall upper selection of $m_{\ell\ell} < 5$ TeV.
In fact, since the bulk of the 1 $b$-tagged jet events lie below $m_{\ell\ell}\simeq 2.5$ TeV (see Fig.~\ref{fig:inclusive_1b_observables}), the EFT parameter 
is $\mathcal{R}_\Lambda \simeq 1$ also if the scale of the NP is 
$\Lambda \gsim 2-3$ TeV (i.e., corresponding to the case $g=1$), so that
the EFT validity is borderline in this case.
Note that 
a value of $\mathcal{R}_\Lambda \sim 0.25$ would correspond in our case to 
a NP scale of $\Lambda \simeq 4-6$ TeV, which is the expected bound range 
in the 1 $b$-tagged jet selection for an intermediate coupling in the underlying heavy physics, i.e., 
$g \sim 2$. 
Similar arguments apply also for the other selections.


\section{Summary}
\label{sec:conclusion}

We have used EFT techniques to carefully analyze the LHC di-lepton TeV scale spectrum in the presence of new $b \bar{b} \ell^+ \ell^-$ contact interactions with a typical scale of $\Lambda \sim {\cal O}(1-10)$ TeV. 
We have considered pair production of either electrons or muons in association with $b$-quarks and studied the high energy behavior of these new interactions.
We find that this form of NP is highly sensitive to the $b$-jet multiplicity in the final state, and that a selection of a single $b$-tagged jet allows to extract improved bounds compared to prior constraints from LEP and to constraints from non-resonant LHC di-lepton searches.
For example, applying an exclusive 1 $b$-tagged jet selection on the inclusive di-electron (di-muon) sample extends the reach on the scale of the $b \bar{b} \ell^+ \ell^-$ operator in the strongly coupled NP case from $\Lambda \sim 5.5\,(5.6)$ to $\Lambda \sim 7.6\,(7.8)$ TeV, assuming 25\% uncertainty for the SM background.
We show that a similar increase of sensitivity is also expected in the weakly interacting underlying heavy physics case, for which a study of the EFT validity is presented.

\acknowledgments
We thank Gauthier Durieux for a useful discussion regarding the choice of the flavor scheme.
This research was supported by a grant from the United States-Israel Binational Science Foundation (BSF) (grant number 2016117), Jerusalem, Israel, and by a grant from the Israel Science Foundation (ISF) (grant number 2871/19).



\bibliographystyle{hunsrt.bst}
\bibliography{mybib2}

\end{document}